\documentclass[aps,prl,floatfix,onecolumn]{revtex4-1}
\usepackage{times}
\usepackage{graphicx}
\usepackage{amsmath}
\usepackage{amssymb}
\usepackage{textcomp} 
\usepackage{bm}
\usepackage{dcolumn}
\usepackage{colortbl}

\newcommand{\bs}[1]{{\boldsymbol{#1}}}
\newcommand{\ie}{{\text{i.e.,}}}

\begin{document}
\title{Generation of a Bessel beam in FDTD using a cylindrical antenna}

\author{Kazem Ardaneh(*), Remo Giust, Benoit Morel and Francois Courvoisier($\dagger$)\\
FEMTO-ST institute, Univ. Bourgogne Franche-Comt\'e, CNRS,\\
15B avenue des Montboucons, 25030, Besan\c{c}on Cedex, France\\
(*)kazem.arrdaneh@gmail.com\\
($\dagger$)francois.courvoisier@femto-st.fr\\
\vspace{2cm}
This is a post-peer-review, pre-copyedit version of an article published in Optics Express. The final authenticated version is available online at: \\
\url{https://doi.org/10.1364/OE.385413} \\
\vspace{2cm}
}

\begin{abstract}
 Bessel beams are becoming a very useful tool in many areas of optics and photonics, because of the invariance of their intensity profile over an extended propagation range. Finite-Difference-Time-Domain (FDTD) approach is widely used for the modeling of the beam interaction with nanostructures. However, the  generation of the Bessel beam in this approach is a computationally challenging problem. In this work, we report an approach for the generation of the infinite Bessel beams in three-dimensional FDTD. {{It is based on the injection of the Bessel solutions of Maxwell's equations from a cylindrical hollow annulus. This configuration is compatible with Particle In Cell simulations of laser plasma interactions.}} This configuration allows using a smaller computation box and is therefore computationally more efficient than the creation of a Bessel-Gauss beam from a wall and models more precisely the analytical infinite Bessel beam. Zeroth and higher-order Bessel beams with different cone angles are successfully produced. We investigate the effects of the injector parameters on the error with respect to the analytical solution. In all cases, the relative deviation is in the range of 0.01-7.0 percent.
\end{abstract}
\maketitle


\section{Introduction}\label{Introduction}
A Bessel beam refers to a family of solutions for the wave equation in which the amplitude of field is expressed by the Bessel function of the first kind \cite{Durnin87}. It has attracted great interest in various branches in optics because  its intensity profile is invariant as it propagates. Bessel beams are mainly utilized as optical traps \cite{Little04,Mcleod08,Vetter19}, for optical manipulation \cite{Garc2002}, optical acceleration \cite{Hafizi97,Li05,Kumar2017}, light-sheet microscopy \cite{Fahrbach2010,Planchon2011}, nonlinear optics, ultrashort pulse filamentation \cite{Gaizauskas06,Roskey07,Faccio12}, and laser-material processing \cite{Duocastella12,Xie2015,COURVOISIER2016,Razvan2018,Bergner18}.
Our main interest here is that Bessel beams can be used to generate long plasma rods with quasi-uniform density \cite{Durfee1993,Hine16}.

Particle-In-Cell (PIC) simulation \cite{Dawson83,bir91} is a conventional pathway to self-consistently model the interaction between laser beams with plasmas. It uses the Finite-Difference-Time-Domain (FDTD) algorithm to advance the electric and magnetic fields. A specific challenge arises for the simulation of Bessel beams in FDTD because of their high aspect ratio. Bessel beams are an interference field in which the propagation length $Z_{\rm B}$ is defined by the transverse extent $W$ and the cone angle $\theta$, {\it i.e.} the angle made by the interfering waves with the optical axis: $Z_{\rm B} \sim W/\tan \theta$ \cite{McLeod54,McGloin05}. Hence, long propagation distances require a wide transverse extent. However, the central spot radius ($r_0=0.383\lambda/\sin\theta$ for a zeroth-order Bessel beam) is generally much smaller than the beam transverse extent. It is therefore computationally extremely demanding to investigate the laser-plasma interaction within the central lobe with high spatial resolution. This problem is illustrated in Fig. \ref{bg}, where we show the time-integrated fluence distribution for a finite-energy zeroth-order Bessel-Gauss femtosecond pulse over a distance of only $\approx20\,\rm{\mu m}$, injected from the left wall of the simulation box. The detail of this simulation is provided in Section \ref{Bessel-Gauss beam}.

Our objective is to replace the simulation box with a smaller one, as shown by transparent white color box in Fig. \ref{bg}. This is possible because in a Bessel beam, energy flows from the sides with a conical structure and because of the longitudinal invariance of both the diffraction-free Bessel beams and the uniform plasma distribution. Therefore, we will use a cylindrical annulus injecting the electromagnetic fields and use periodic boundary condition for the surfaces parallel to the optical axis to reproduce the longitudinal invariance. 

Bessel beams have been previously generated in FDTD using two different approaches.  In \cite{Wu18}, Wu {\it et al} have simulated the generation of Bessel beam sources in FDTD, using total-field/scattered-field method \cite{Taflove2005}. In this approach, the computational domain is split into total-field and scattered-field regions. The electromagnetic wave is injected using the surface which separates the two regions. In \cite{Chen18}, arbitrary order Bessel beams were generated using the scattered-field approach \cite{Taflove2005}. In this case, the total fields are decomposed into known incident fields and unknown scattered fields. The incident fields at all grid points are evaluated using the analytical expression at each time step.

{{Previous techniques of Bessel beam generation have inherent limitations. Indeed}}, both total-field/scattered-field and scattered-field methods are inappropriate for implementation in PIC codes. In the scattered-field approach, there is no direct access to the total field which is required in PIC codes. Moreover, the incident field is needed at any point in the grid. The total-field/scattered-field needs an extra computational region for the scattered field. In this technique particle and field boundaries would be defined in different places because particles are in interaction with the total fields. In contrast with these approaches, injecting the electromagnetic fields using an antenna has several advantages: (1) there is direct access to total field, (2) incident field is needed only at the antenna points, (3) no extra computational region is needed for the scattered fields (because PIC approach entirely works with total field), and (4) particle and field boundaries are set in same place.

In the present work, a cylindrical annulus which we call Bessel antenna is inscribed in the FDTD box and emits the Bessel solutions of Maxwell's equations. In comparison with Bessel-Gauss FDTD simulation, it significantly reduces the size of the three-dimensional (3D) computational box.  It has been successfully tested for different orders of Bessel beam, different cone angles, different antenna thicknesses, and different antenna radii. The relative deviation between the fields from the  Bessel antenna  and those from the analytical solution is in the range of 0.01-7.0 percent. {{We readily note that the more straightforward approach consisting of injecting the Bessel solutions from the computational walls (square symmetry) does not yield satisfactory results in terms of beam symmetry}}.

The paper is organized as follows. We first generate Bessel-Gauss beam in FDTD simulation for reference. Then, we derive in Section \ref{Mathematical model} the Bessel solutions of the Hertz vector potential. We also provide in this section the analytical model of Bessel pulse to which we will compare our numerical simulations. The Bessel antenna implementation is detailed in Section \ref{Bessel antenna implementation}. The results of the simulations are discussed in Section \ref{Results}, where we investigate the influence of the antenna parameters.

\section{Bessel-Gauss beam}\label{Bessel-Gauss beam}

In this section, we implement the generation of a Bessel-Gauss beam in FDTD. Experimentally, a Bessel-Gauss beam can be created by focusing a Gaussian beam using an axicon lens \cite{McLeod54,GORI1987,Jarutis2000,McGloin05}.  The axicon applies a phase $\Phi(r)=-k r\sin\theta$ onto the Gaussian beam, where $r$ is the radial distance in the cylindrical coordinate, $k=2\pi/\lambda$ is the wavenumber with $\lambda$ the laser central wavelength and $\theta$ is the cone angle. 

For this example, we utilize a computational box of $15\times15\times 28\,{\rm{\mu m}}^3$. Using a normalization in terms of wavelength, it can be expressed as $8\lambda_{\rm xy}\times8\lambda_{\rm xy}\times32\lambda_{\rm z}$  where $\lambda_{\rm xy}=\lambda/\sin\theta$, and $\lambda_{\rm z}=\lambda/\cos\theta$. We inject from the $z_{\min}$ wall a Gaussian beam polarized along $x$ direction and propagating in the the positive $z$ direction, on which we apply the cylindrically-symmetric phase $\Phi(r)$. We set the wavelength and cone angle $0.8~{\rm \mu m}$, $25 ^{\circ}$, respectively. The temporal amplitude profile of the beam is a Gaussian function of width $100\,{\rm fs}$ Full Width at Half Maximum (FWHM).  The Gaussian beam waist is $w_0=6\,{\rm\mu m}$.  We run the simulation up to $t_{\rm {run}}=266\,{\rm fs}$.

\begin{figure}[ht]
\begin{center}
\includegraphics[width=0.75\columnwidth]{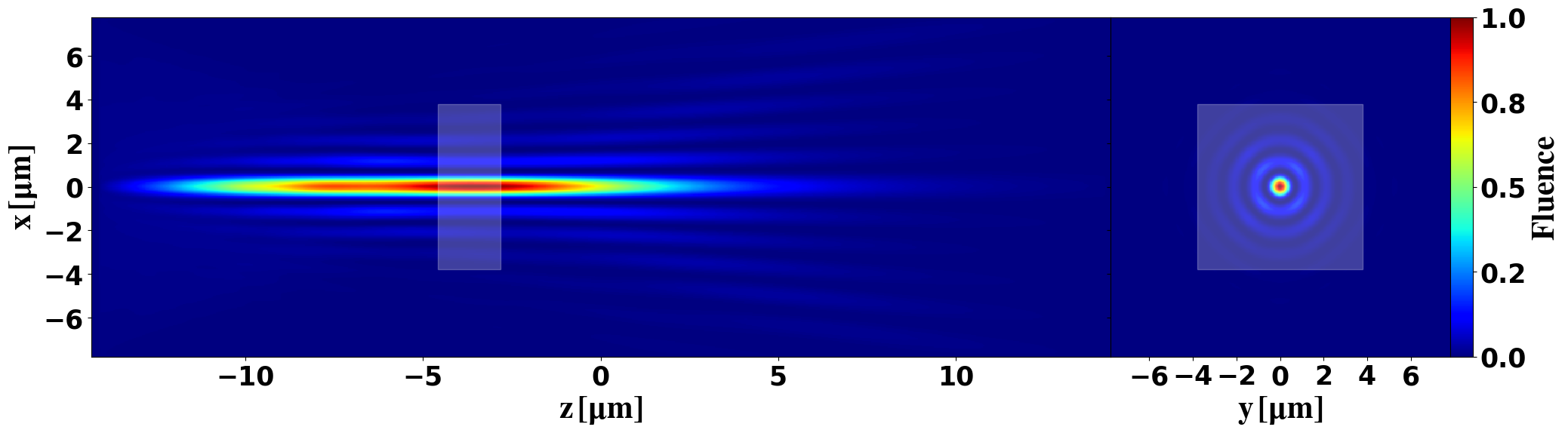}
\caption{A Bessel-Gauss beam from FDTD simulation. Fluence distribution in $zx$ plane at $y=0$ (left), and $yx$ plane at location of the maximum for $I(0,0,z)$ (right). We aim to create a Bessel beam in the white box with a computational box of $4\lambda_{\rm xy}\times4\lambda_{\rm xy}\times2\lambda_{\rm z}$ which is smaller than Bessel-Gauss one by a factor of 64.}
\label{bg}
\end{center}
\end{figure}

The fluence distribution ($\int_{0}^{t_{\rm {run}}}{S~ {\rm d}t}$ where $S$ is the magnitude of the Poynting vector) for the resulting Bessel-Gauss beam at the end of the simulation is shown in Fig. \ref{bg}. As one can see, a Bessel-Gauss beam with a propagation distance of only $\approx 20\,{\rm \mu m}$ already requires a window width of $\approx 15\,{\rm \mu m}$. In this example, the ratio between the central lobe radius and the transverse dimension is about 0.1. 
The interaction of this beam with a nanoscale plasma rod will typically drive electron plasma waves on a spatial scale of $ \sim0.05\lambda$. If we resolve the wavelength of the excited plasma waves with 20 grid cells, a PIC simulation will need a grid of $7500\times 7500\times 14100$ which is computationally very expensive. We have implemented the Bessel antenna as an alternative approach that can create a Bessel beam in a smaller box (transparent white color box in Fig. \ref{bg}).

\section{Mathematical model}\label{Mathematical model}
\subsection{Bessel's solutions}\label{Bessel's solutions}
Using the antenna, we will generate ideal Bessel beams by injecting the solution of the electromagnetic wave equation in cylindrical coordinates.
Several authors have derived the Bessel solutions for Maxwell's equations in a homogeneous, isotropic, and unmagnetized medium \cite{MISHRA1991159,Yu2008VECTORAO,WANG2016218}. But their solutions were not fully general in terms of polarization states. We briefly recall here the general formulas and then provide handy normalized forms. We start with the Hertz vector potential $\bs{\Pi}$ which reads \cite{stratton2007}:

\begin{equation}\label{hertz}
	\bs{\nabla}^2\bs{\Pi}-\mu\,\epsilon\,\partial_t^2\bs{\Pi}=\mathbf{0}
\end{equation}

There are generally two solutions (related to TE and TM modes in cylindrical symmetry) for the electromagnetic fields satisfying Eq. (\ref {hertz}). These solutions take the following forms for a harmonic function as $\exp(-i\omega t)$:

\begin{subequations}
\begin{eqnarray}
	(\mathbf{E}_1,\mathbf{E}_2)&=&(\bs{\nabla}\times\bs{\nabla}\times\bs{\Pi},i\,\omega\,\mu\,\bs{\nabla}\times\bs{\Pi})\\
	(\mathbf{H}_1,\mathbf{H}_2)&=&(-i\,\omega\,\epsilon\,\bs{\nabla}\times\bs{\Pi},\bs{\nabla}\times\bs{\nabla}\times\bs{\Pi})
\end{eqnarray}
\end{subequations}

The solutions can be determined using two functions given by:

\begin{equation}\label{QR}
	(\mathbf{Q},\mathbf{R})=(\bs{\nabla}\times\bs{\Pi},\bs{\nabla}\times\bs{\nabla}\times\bs{\Pi})
\end{equation}

Hence:

\begin{subequations}
\begin{eqnarray}
\label{E1E2}
	(\mathbf{E}_1,\mathbf{E}_2)&=&(\mathbf{R},i\,k_0\,\mathbf{Q})\\
	(\eta\,\mathbf{H}_1,\eta\,\mathbf{H}_2)&=&(-i\,k_0\,\epsilon_{\rm r}\,\mathbf{Q},\mathbf{R})
\end{eqnarray}
\end{subequations}

here $k_0=\omega/c$ denotes the wavenumber in the vacuum, $\eta=\mu_0\,c$ the impedance of free space, and $\epsilon_{\rm r}$ the relative permittivity of the propagating medium. For Bessel's solutions of Eq. (\ref{hertz}), one of the following Hertz potentials is usually employed \cite{stratton2007,MISHRA1991159,Yu2008VECTORAO,WANG2016218}.

\begin{subequations}\label {solutions}
\begin{eqnarray}
	\bs{\Pi}_{\rm {x}}&=& {\mathbf{\hat{x}}} J_{\rm m}(K\,r)e^{i\,m\,\phi}\,e^{i(\omega\,t-\beta\,z)}\\
	\bs{\Pi}_{\rm {y}}&=& {\mathbf{\hat{y}}} J_{\rm m}(K\,r)e^{i\,m\,\phi}\,e^{i(\omega\,t-\beta\,z)}\\
	\bs{\Pi}_{\rm {z}}&=& {\mathbf{\hat{z}}} J_{\rm m}(K\,r)e^{i\,m\,\phi}\,e^{i(\omega\,t-\beta\,z)}
\end{eqnarray}
\end{subequations}

where $K$ is the transverse component (in $xy$ plane) of the wave vector, $\beta$ the axial component (along the $z$ axis) of the wave vector,  $m$ order of Bessel function, $\phi$ the azimuth angle, and $(\mathbf{\hat{x},\hat{y},\hat{z}})$ are unit vectors of cartesian coordinates. We have, therefore, $\beta^2+K^2=\epsilon_{\rm r}\,k_0^2 = k^2$. To link with the cone angle $\theta$ described in the previous section, we have $K=k \sin \theta$ and $\beta = k \cos \theta$.

Substituting Eq. (\ref{solutions}) into Eq. (\ref{QR}) and after some algebra, we can obtain $(\mathbf{Q},\mathbf{R})$ vectors associated with the three potentials $(\bs{\Pi}_{\rm {x}},\bs{\Pi}_{\rm {y}},\bs{\Pi}_{\rm {z}})$. In Cartesian coordinates, they are given by:

\begin{enumerate}
\item{$\bs{\Pi}_{\rm x}$}

\begin{subequations}
\begin{eqnarray}
	\mathbf{Q}&=&e^{i(\omega\,t-\beta\,z)}\,e^{i\,m\,\phi}\,
	\begin{array}{|l}
		\mathbf{\hat{x}}\,0 \\ 
		\mathbf{\hat{y}}\,i\,\beta\,J_{\rm m}(K\,r) \\ 
		\mathbf{\hat{z}}\,\frac{K}{2\,i}\left[J_{\rm {m+1}}(K\,r)\,e^{i\,\phi}+J_{\rm {m-1}}(K\,r)\,e^{-i\,\phi}\right] \\ 
	\end{array}\\ 
	\mathbf{R}&=&e^{i(\omega\,t-\beta\,z)}\,e^{i\,m\,\phi}\,
	\begin{array}{|l}
		\mathbf{\hat{x}}\,\frac{k^2+\beta^2}{2}\,J_{\rm m}(K\,r)+
		\frac{K^2}{4}\,\left[
		J_{\rm {m+2}}(K\,r)\,e^{2\,i\,\phi}+J_{\rm {m-2}}(K\,r)\,e^{-2\,i\,\phi}\right]\\ 
		\mathbf{\hat{y}}\,i\,\frac{K^2}{4}\,\left[
		J_{\rm {m-2}}(K\,r)\,e^{-2\,i\,\phi}-J_{\rm {m+2}}(K\,r)\,e^{2\,i\,\phi}\right]\\ 
		\mathbf{\hat{z}}\,\frac{i\,\beta\,K}{2}\left[J_{\rm {m-1}}(K\,r)\,e^{-i\,\phi}-J_{\rm {m+1}}(K\,r)\,e^{i\,\phi}\right] \\ 
	\end{array}
\end{eqnarray}
\end{subequations}

\item{$\bs{\Pi}_{\rm y}$}
\begin{subequations}\label{Py}
\begin{eqnarray}
	\mathbf{Q}&=&e^{i(\omega\,t-\beta\,z)}\,e^{i\,m\,\phi}\,
	\begin{array}{|l}
		-\mathbf{\hat{x}}\,i\,\beta\,J_{\rm m}(K\,r) \\ 
		\mathbf{\hat{y}}\,0 \\ 
		-\mathbf{\hat{z}}\,\frac{K}{2}\left[J_{\rm {m+1}}(K\,r)\,e^{i\,\phi}-J_{\rm {m-1}}(K\,r)\,e^{-i\,\phi}\right] \\ 
	\end{array}\\ 
	\mathbf{R}&=&e^{i(\omega\,t-\beta\,z)}\,e^{i\,m\,\phi}\,
	\begin{array}{|l}
		\mathbf{\hat{x}}\,i\,\frac{K^2}{4}\,\left[
		J_{\rm {m-2}}(K\,r)\,e^{-2\,i\,\phi}-J_{\rm {m+2}}(K\,r)\,e^{2\,i\,\phi}\right]\\ 
		\mathbf{\hat{y}}\,\frac{k^2+\beta^2}{2}\,J_{\rm m}(K\,r)-
		\frac{K^2}{4}\,\left[
		J_{\rm {m+2}}(K\,r)\,e^{2\,i\,\phi}+J_{\rm {m-2}}(K\,r)\,e^{-2\,i\,\phi}\right]\\ 
		-\mathbf{\hat{z}}\,\frac{\beta\,K}{2}\left[J_{\rm {m+1}}(K\,r)\,e^{i\,\phi}+J_{\rm {m-1}}(K\,r)\,e^{-i\,\phi}\right] \\ 
	\end{array}
\end{eqnarray}
\end{subequations}

\item{$\bs{\Pi}_{\rm z}$}
\begin{subequations}\label{Pz}
\begin{eqnarray}
	\mathbf{Q}&=&e^{i(\omega\,t-\beta\,z)}\,e^{i\,m\,\phi}\,
	\begin{array}{|l}
		\mathbf{\hat{x}}\,i\,\frac{K}{2}\left[J_{\rm {m+1}}(K\,r)\,e^{i\,\phi}+J_{\rm {m-1}}(K\,r)\,e^{-i\,\phi}\right]\\ 
		\mathbf{\hat{y}}\,\frac{K}{2}\left[J_{\rm {m+1}}(K\,r)\,e^{i\,\phi}-J_{\rm {m-1}}(K\,r)\,e^{-i\,\phi}\right]\\ 
		\mathbf{\hat{z}}\,0\\ 
	\end{array}\\ 
	\mathbf{R}&=&e^{i(\omega\,t-\beta\,z)}\,e^{i\,m\,\phi}\,
	\begin{array}{|l}
		-\mathbf{\hat{x}}\,i\,\frac{\beta\,K}{2}\left[J_{\rm {m+1}}(K\,r)\,e^{i\,\phi}-J_{\rm {m-1}}(K\,r)\,e^{-i\,\phi}\right]\\ 
		-\mathbf{\hat{y}}\,\frac{\beta\,K}{2}\left[J_{\rm {m+1}}(K\,r)\,e^{i\,\phi}+J_{\rm {m-1}}(K\,r)\,e^{-i\,\phi}\right]\\ 
		\mathbf{\hat{z}}\,K^2\,J_{\rm m}(K\,r)\\ 
	\end{array}
\end{eqnarray}
\end{subequations}
\end{enumerate}

The first two solutions, \ie $\bs{\Pi}_{\rm x}$ and $\bs{\Pi}_{\rm y}$, specify Bessel beams with linear, and more generally elliptical polarizations (because the field has in general a $z$ component), while the solution $\mathbf{\Pi_{\rm z}}$ generate the purely radial or azimuthal polarizations for $m=0$ (using respectively $\mathbf{E}_1$ and $\mathbf{E}_2$ solutions of Eq. (\ref{E1E2})).
One can exactly reproduce Eqs. (7), and (8) in \cite{WANG2016218} from $\bs{\Pi}_{\rm y}$ in Eqs. (\ref{Py}) and Eqs. (12a) and (12b) in \cite{Yu2008VECTORAO} from $\bs{\Pi}_{\rm z}$ in Eqs. (\ref{Pz}).  

{{For implementation, we modify the above solutions based on cone angle $\theta$ and the amplitude of the electric field $E_0$. We note that $E_0$ corresponds, in the case $m=0$, to the modulus of the electric field on the optical axis ($r=0$). The corresponding $(\mathbf{E},\mathbf{H})$ fields for TE, and TM modes are given in Eq. (\ref {TETM}a), and Eq. (\ref {TETM}b), respectively.}}

\begin{subequations}\label {TETM}
\begin{eqnarray}
		(\mathbf{E},\mathbf{H})&=&k_0^{-2}(E_0k_0\mathbf{Q}, H_0\mathbf{R})\,e^{i(\omega\,t-\beta\,z)}\,e^{i\,m\,\phi}\\
		(\mathbf{E},\mathbf{H})&=&k_0^{-2}(E_0\mathbf{R},H_0k_0\mathbf{Q})\,e^{i(\omega\,t-\beta\,z)}\,e^{i\,m\,\phi}\
\end{eqnarray}
\end{subequations}


{{where $H_0={n}/{\eta}\,E_0$ (in terms of the energy density $\mu_0H_0^2=\epsilon_0\,\epsilon_{\rm r}\,E_0^2$), and  $n=\epsilon_{\rm r}^{1/2}$ is the optical refractive index.}}

\begin{figure}[h!]
\begin{center}
\includegraphics[width=0.75\columnwidth]{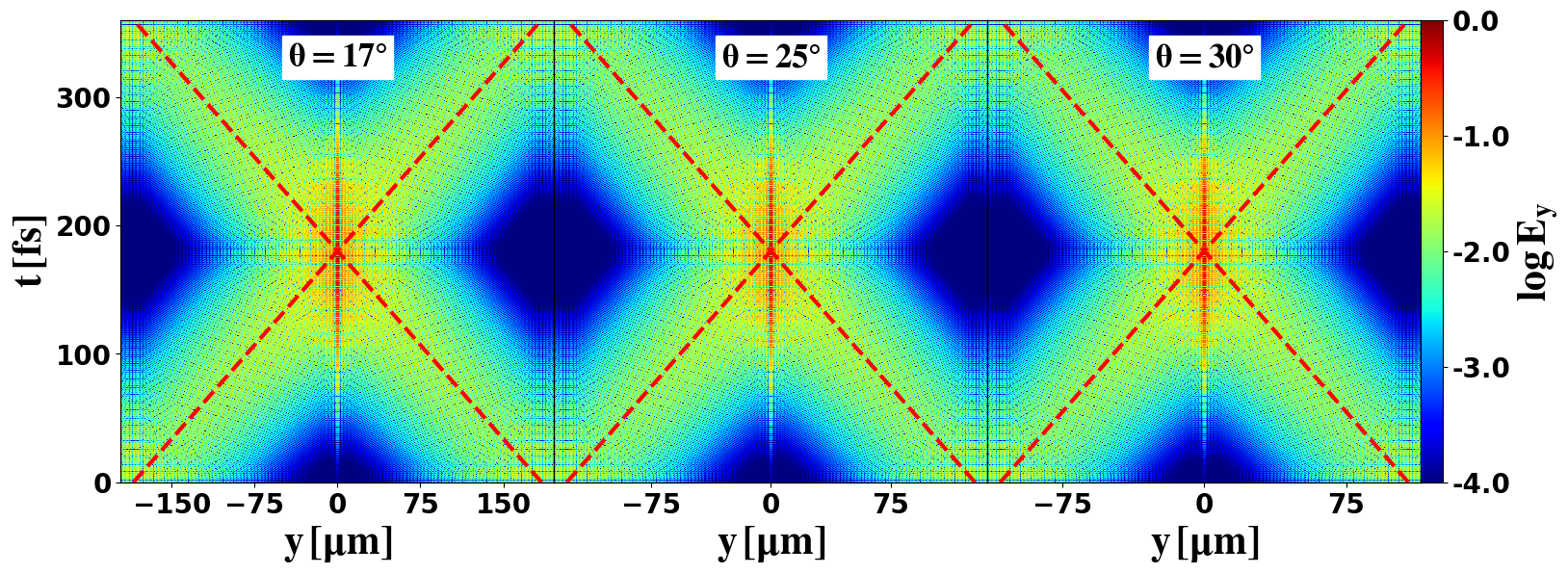}
\caption{Bessel pulse from Fourier spectrum integration. Eq. (\ref{pulse}) integration is calculated for cone angle of $\theta=17 ^{\circ}$ (left panel), $\theta=25 ^{\circ}$ (middle panel), and $\theta=30 ^{\circ}$ (right panel). In all cases, $m=0$, $T=60\,{\rm fs}$, $T_0=180\,{\rm fs}$, and $\omega_0=2.4\,{\rm{PHz}}$. The red dashed lines show $\pm c/\sin\theta$ velocities. We show the amplitude of $E_{\rm y}$ field in log-scale to enhance the X-shape of the pulse propagation.}
\label{pulsef}
\end{center}
\end{figure}
\subsection{Bessel pulse}\label{Bessel pulse}
For the validation of the antenna-generated Bessel pulse, we will need a reference based on an analytical expression. Here, we express the Bessel pulse using the monochromatic field derived in section \ref{Bessel's solutions}. We work hereafter with the vector potential polarized along the $x$ axis ($\bs{\Pi}_{\rm {x}}$) to generate a linearly $y$ polarized Bessel pulse of arbitrary order. A pulse in time-domain is obtained by performing the integral over the pulse spectrum $F(\omega)$. We choose  $F(\omega)$ as a Gaussian function that inverse Fourier transform gives $F(t)=\exp(-i\omega_0 t)\exp(-t^2/T^2)$ function in time domain. Hence:

\begin{equation}\label{guass}
F(\omega)=\frac{T}{2\sqrt{\pi} }e^{-\frac{T^{2}(\omega-\omega_{0})^{2}}{4}}
\end{equation}

\noindent  The $y$ component of the electric field then reads:

\begin{equation}\label{pulse}
E_{\rm y}(\mathbf{x},t)= -\cos\theta\,e^{i\,m\,\phi}\int_{-\infty}^{\infty}{d\omega F(\omega)J_{\rm m}(\frac{\omega\sin{\theta}}{c}r)e^{i\omega(t-T_{0}-z\cos\theta/c)}} 
\end{equation}

Where $T_0$ corresponds to the time of the peak intensity. 

We calculate the Eq. (\ref{pulse}) integration for $m=0,1,2$, $T=60\,{\rm fs}$ (${\rm FWHM}=2T\sqrt{\ln{2}}=100\,{\rm fs}$), $T_0=180\,{\rm fs}$, and $\omega_0=2.4\,{\rm{PHz}}$. Three different cone angles of $\theta=17 ^{\circ}$, $\theta=25 ^{\circ}$, and $\theta=30 ^{\circ}$ are considered. Shown in Fig. \ref{pulsef} is the absolute value of $E_{\rm y}$ (normalized to the maximum value) extracted from Eq. (\ref{pulse}) for $m=0$. The $y$ coordinate is scaled with the transverse wavelength $\lambda_{\rm xy}=\lambda/\sin\theta$. It, therefore, increases as the cone angle decreases from $30 ^{\circ}$ to $17^{\circ}$. In all cases, one sees two propagating waves coming from the bottom ($t=0$), interfere around $y=0$, and then propagate away. The interfering waves propagate with velocities of $\pm c/\sin\theta$ (red dashed lines). In Section \ref{Results}, the $E_{\rm y}$ of the antenna generated Bessel pulse will be compared with the corresponding Eq. (\ref{pulse}) integration. We note again that the width over which we see the pulse propagation is much larger than the width of the central lobe.

\section{Bessel antenna implementation}\label{Bessel antenna implementation}
To model the propagation of a plane wave in FDTD, we inject ${\bf{E}}$ and ${\bf{B}}$ solutions of the wave equation from one wall into the computational box. We use a similar methodology for the propagation of the Bessel beam. Since a Bessel beam has a cylindrical symmetry, we inject the Bessel solution presented in Section \ref{Bessel's solutions} from a cylindrical antenna into the simulation box. 

\begin{figure}
\begin{center}
\includegraphics[width=0.75\columnwidth]{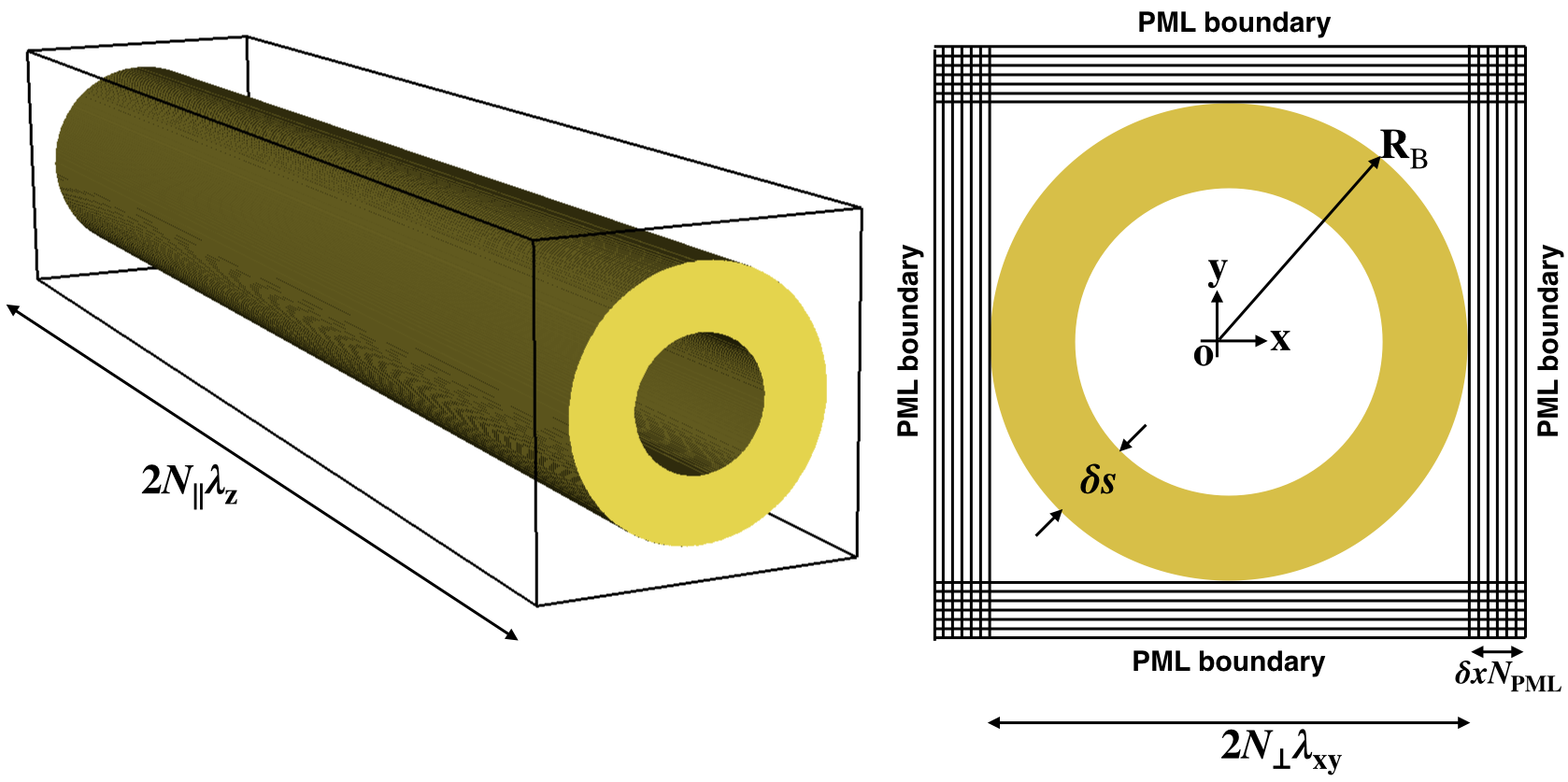}
\caption{Concept of the Bessel antenna in the FDTD simulation box. The yellow annulus with a thickness of $\delta s$ shows the cylindrical antenna which has an outer radius $R_{\rm B}=N_{\perp}\lambda_{\rm xy}$, and a length $2N_{\parallel}\lambda_{\rm z}$. There are $N_{\rm PML}$ grid cells in each transverse direction where the CPML boundary condition is applied.}
\label{PML}
\end{center}
\end{figure}

We implement the antenna in EPOCH code, which is a Particle-In-Cell (PIC) code for laser-plasma simulation \cite{Arber_2015}. The FDTD algorithm in EPOCH uses a modified version of the leapfrog scheme in which the fields are updated at both the half time-step and the full time-step. The full detail of the EPOCH code is presented in \cite{Arber_2015}. The standard second order Yee's FDTD scheme is used in the present work, although higher-order schemes (4th- and 6th-order) are also available.

The  Bessel antenna is inscribed in the simulation box as illustrated in Fig. \ref{PML}. The cylinder has an outer radius $R_{\rm B}$ and a thickness $\delta s$. 
To avoid reflection of the outgoing waves from the $(x_{\min},x_{\max})$ and $(y_{\min},y_{\max})$ boundaries, perfectly matched layers (PML) are used in these surfaces as shown in Fig. \ref{PML}. A PML is an artificial absorbing layer. It is commonly used to truncate computational regions in numerical methods to simulate problems with open boundaries, particularly in the FDTD \cite{Taflove2005}. In practice, we used the convolutional PML (CPML) method, as presented in \cite{Taflove2005,Roden2000}. 
The current CPML implementation in EPOCH uses a number of PML layers ranging between  6 and 16. We use $N_{\rm{PML}}=6$ in the current work.

\begin{table*}
\begin{center}
\caption{Sets of simulation parameters for Bessel antenna.\label{parameters}}
\begin{tabular}{ |c||c||c||c||c||c||c||c||c||c|  }
 \hline
Run& A & B & C & D & E & F & G & H & I\\
 \hline
$m$& 0 & 1 & 2 & 0 & 0 & 0 & 0 & 0 & 0 \\
$\theta^{\circ}$& 30 & 30 & 30 & 25 & 17 & 25 & 25 & 25 & 25 \\
$\delta s\,[\lambda]$& 1 & 1 & 1 & 1 & 1 & 2 & 3 & 1 & 1\\
$N_{\parallel}$& 1 & 1 & 1 & 1 & 1 & 1 & 1 & 1 & 1\\
$N_{\perp}$& 4 & 4 & 4 & 4 & 4 & 4 & 4 & 3 & 2\\
 \hline
\end{tabular}
\end{center}
\end{table*}

We define the sampling of our computational box with the following procedure. We sample the laser wavelength with $n_{\rm s}$ grid cells. Here, we use $n_{\rm s}=20$.  We set the transverse width of the box, excluding the PML boundary layers, as an even number $2N_{\perp}$ of the transverse period of the field $\lambda_{\rm xy} =2\pi/K$. The number of points of the computational box is then defined as $N_{\rm xy}={\rm{int}}(2n_{\rm s}N_{\perp}/\sin\theta)+2N_{\rm{PML}}$ in the transverse plane. The longitudinal size of the box, including the PML boundary layers, is an even number of the period $\lambda_{\rm z} =2\pi/\beta$  and the corresponding number of points is $N_{\rm z}={\rm{int}}(2n_{\rm s}N_{\parallel}/\cos\theta)+2N_{\rm{PML}}$. This ensures that the wave propagation is accurately sampled in both longitudinal and transverse directions. The cell size is also same in both directions so that the mesh is cubic. (We note, for EPOCH users, that in EPOCH input parameter file, the number of grid points and length of the computational box have to be reduced by the number of CPML layers and corresponding width of the CPML conditions.)

We will focus on the generation of linearly polarized Bessel beams of order $m$ with a Gaussian temporal profile. Hence, we use the vector potential polarized along the $x$ axis. As shown in section \ref{Bessel's solutions}, this will generate a linearly $y$ polarized Bessel beam.  Since the Fourier transformation operation cannot be straightforwardly implemented in EPOCH, we use the zeroth-order approximation to the integration of Eq. (\ref{pulse}), which corresponds to multiplying the $({\bf{E}},{\bf{B}})$ solutions of Eq. (\ref{TETM}a) or Eq. (\ref{TETM}b) a Gaussian time profile peaked at time $T_0$: $f(t)=\exp[{{-(t-T_{0})^2}/{T^2}}]$. Therefore, our input fields are defined by: 

\begin{subequations}\label {bessel_a} 
\begin{eqnarray}
		\mathbf{E}_{\rm B}&=&k_0^{-2}\Re\{E_0f(t)k_0\mathbf{Q}\,e^{i[\omega\,(t-T_0)-\beta\,z]}\,e^{i\,m\,\phi}\}\\
		\mathbf{H}_{\rm B}&=&k_0^{-2}\Re\{H_0f(t)\mathbf{R}\,e^{i[\omega\,(t-T_0)-\beta\,z]}\,e^{i\,m\,\phi}\}\
\end{eqnarray}
\end{subequations}
	
Here we note that the zeroth-order approximation we performed is equivalent to neglecting the time needed by an illuminating temporally Gaussian pulse to reach the extremity of the cylinder in front of the pulse duration. In other words, we consider that the Gaussian pulse illuminates at the same time the front side ($z_{\rm min}$) and the rear side ($z_{\rm max}$). This is necessary to obtain the periodic boundary conditions between these two planes. This approximation is fully justified if the computational box size is much smaller than the pulse duration divided by the speed of light.

	\begin{figure}[hbtp]
\begin{center}
\includegraphics[width=0.75\columnwidth]{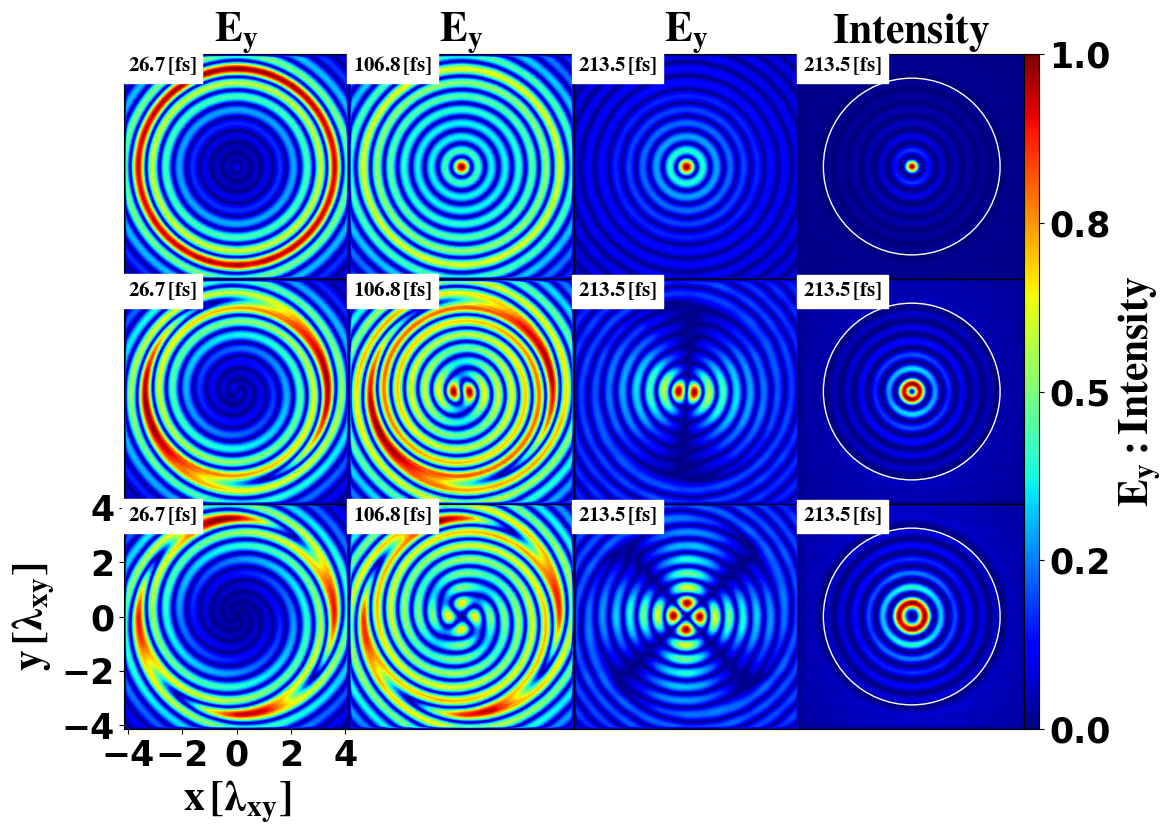}
\caption{Different-order Bessel beams. The amplitude of the $y$ component electric field and the intensity of generated Bessel beams are plotted on $xy$ plane at $z=0$: Run A (top row),  Run B (middle row), and Run C (bottom row). The first three columns show the $y$ component of the electric field at different times, $t=26.7\, [{\rm fs}]$, $106.8\, [{\rm fs}]$, and $213.5\, [{\rm fs}]$ from left to right. The rightmost column shows the intensity distribution at $t=213.5\, [{\rm fs}]$, the corresponding time of the peak intensity ($T_{0}+R_{\rm B}/c\sin\theta$). The white circle in the right column shows the inner radius of the  Bessel antenna .}
\label{diffm}
\end{center}
\end{figure}

Finally, we need to define the procedure with which the fields will be updated at the antenna points.
In the common source implementation, which is referred to as hard source \cite{Taflove2005}, the fields are defined in source points from predefined functions (like Eqs. \ref{bessel_a}) and the FDTD update equations do not apply to them. This kind of source implementation, however, scatters all incident waves on the source. In our case, however, the fields emitted out of a plasma placed inside the cylindrical antenna should not bounce back. 
 
Therefore, we use the soft source scheme \cite{Schneider98} which drastically reduces the scattering of the incoming waves.  We update the points in the antenna by the sum of the values from the FDTD Maxwell's curl equations and the values dictated by the Eqs. \ref {bessel_a}. Hence 

\begin{subequations}\label{source}
\begin{align}
E^{\rm {t+\delta t}}_{\rm i}(i+\delta_{\rm i}/2, j, k)=&E^{\rm {t+\delta t}}_{\rm i}(i+\delta_{\rm i}/2, j, k) + E^{\rm {t+\delta t}}_{\rm Bi}(i+\delta_{\rm i}/2, j, k)\\
\begin{split}
H^{\rm {t+\delta t}}_{\rm i}(i, j+\delta_{\rm j}/2, k+\delta_{\rm k}/2)=&H^{\rm {t+\delta t}}_{\rm i}(i, j+\delta_{\rm j}/2, k+\delta_{\rm k}/2)+ \\
&H^{\rm {t+\delta t}}_{\rm Bi}(i, j+\delta_{\rm j}/2, k+\delta_{\rm k}/2)
\end{split}
\end{align}
\end{subequations}

with cyclic permutation of the indices $i$, $j$, and $k$ corresponding to $x$, $y$, and $z$, respectively. Here, $\delta_i$ is the cell size in direction $i$.

We note that we have independently investigated the opportunity of implementing a transparent source \cite{Schneider98}, which is based on subtracting impulse response of the grid. However, given the high sampling of our grid ($n_{\rm s}=20$ grid cells per wavelength) and introducing the source with a smoothly increasing envelope, it appeared that the difference between the soft source and the transparent one would be less than 5 percent. In contrast, the calculation of the convolution of the injected fields with the impulse response at all times would be computationally very expensive, this is why we discarded this technical solution. All results presented afterwards were obtained using the soft source defined in Eq. (\ref{source}).

\section{Results}\label{Results}

In this section, we present our results and compare them to the analytical one from Eq. (\ref{pulse}). The antenna is tested for Bessel order of $m=0$, $1$, and $2$, and the cone angle of $\theta=17 ^{\circ}$, $\theta=25 ^{\circ}$, and $\theta=30 ^{\circ}$. The different sets of parameters are summarized in Table \ref{parameters}. For all cases, $T=60\,{\rm fs}$, $T_0=180\,{\rm fs}$, and $\lambda=0.8\,{\rm \mu m}$.  

We show in Fig. \ref{diffm} in the three leftmost columns the amplitude of $E_{\rm y}$ in $xy$ plane at different times in the pulse for the FDTD generated Bessel beams of order $m=0$, $1$, and $2$.  These are calculated for a cone angle of $\theta=30 ^{\circ}$. The rows correspond respectively to runs A, B, C where the order $m$ was varied.
We observe that the cylindrical symmetry is well preserved for $m=0$ and that during the pulse build-up, the fields propagate inwards. The interference progressively takes place at the center to create the Bessel beams of order $m$. For the cases of $m=1,2$, the pattern effectively rotates with time. (We remark that maxima have the same orientation in the 3 times because they are separated by integer number of temporal periods.)

In the rightmost column of Fig.\ref{diffm}, we show the intensity distribution at a time $t=213.5~{\rm fs}$, {\it i.e.} which corresponds to the time of the peak intensity. The intensity is calculated using $1/\tau\int_{t}^{t+\tau}{dt\,S}$ where $\tau$ is the period and $S$ is the magnitude of the Poynting vector. The intensity distribution shows the expected transverse map for the different order Bessel beams. There is a high-intensity core in the zeroth-order Bessel beam while one can see the low-intensity core for the first-order and the second-order beams. The cylindrical symmetry is highly apparent, confirming the quality of the injection. 

\begin{figure}[hbtp]
\begin{center}
\includegraphics[width=0.75\columnwidth]{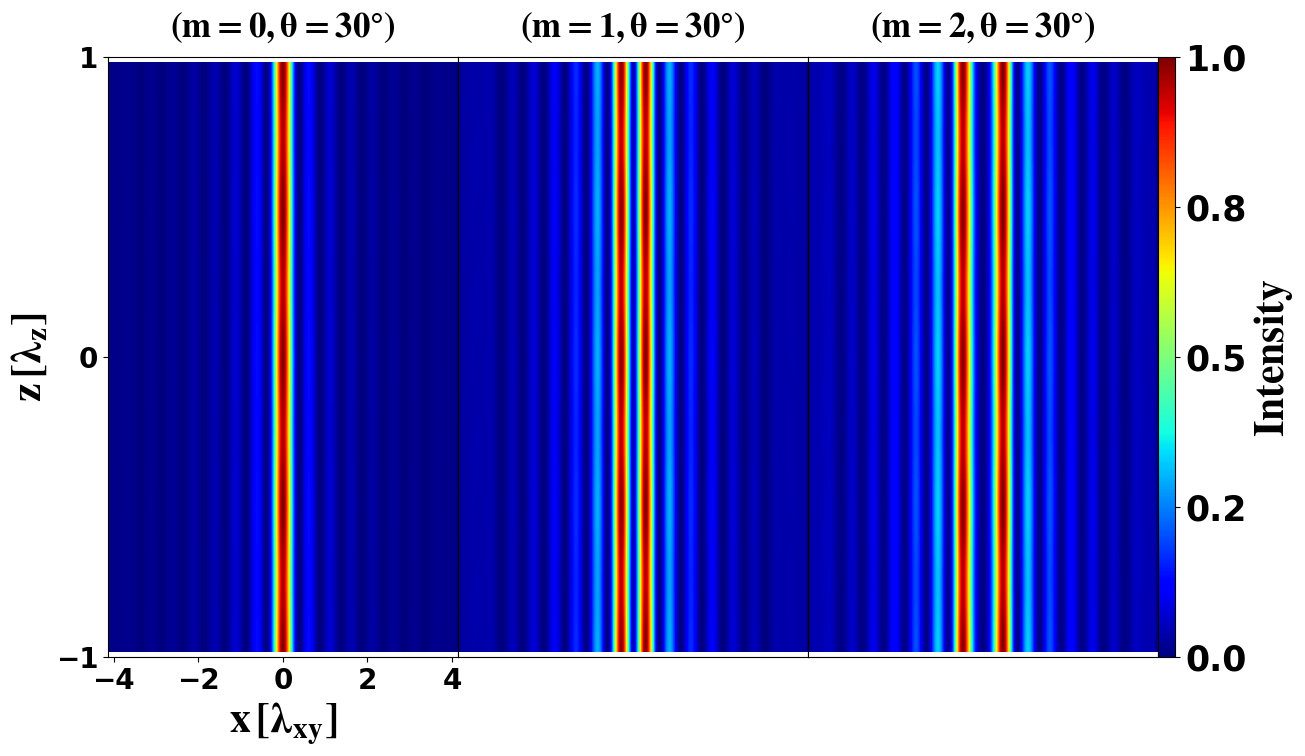}
\caption{Different-order Bessel beams. The intensity map of generated Bessel beams in $xz$ plane at $y=0$: Run A (left),  Run B (middle), and Run C (right). All snapshots are taken at $213.5\, {\rm fs}$.}
\label{diffm2}
\end{center}
\end{figure}

In Fig. \ref{diffm2} we show the intensity distribution in $xz$ plane at a time $t=213.5~{\rm fs}$. The intensity is actually invariant with $z$ within $\pm 2$ percent, confirming the "non-diffracting" behaviour.
 
\begin{figure}[hbtp]
\begin{center}
\includegraphics[width=0.75\columnwidth]{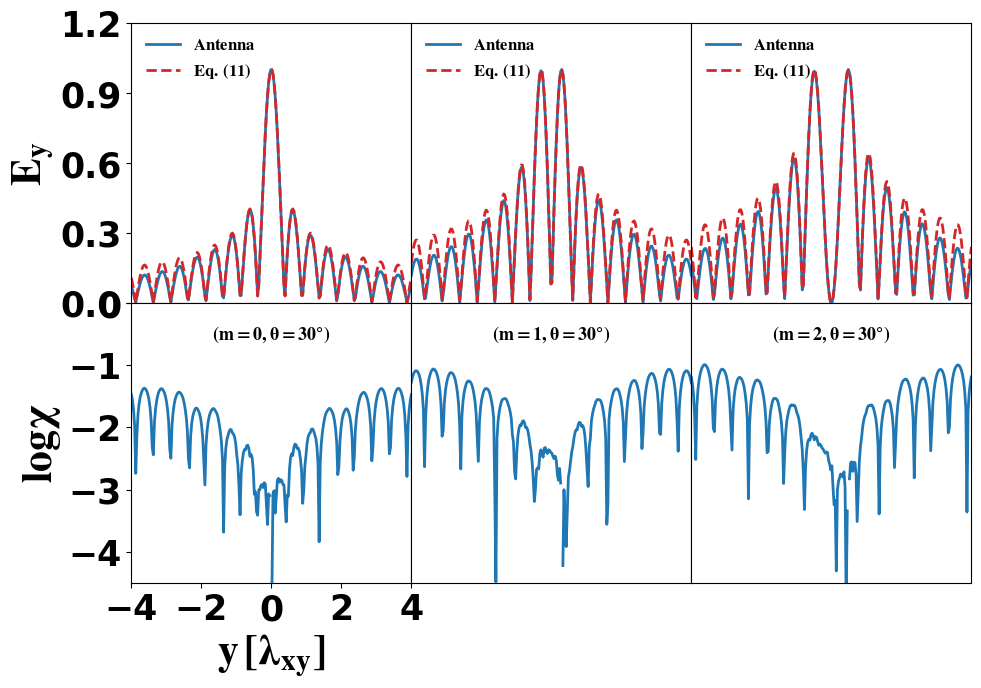}
\caption{Different-order Bessel beams. Comparisons between the $y$ component electric field from the Bessel antenna (blue solid line) and from the Eq. (\ref{pulse}) (red dashed line). Run A (left column),  Run B (middle column), and Run C (right column). The bottom row shows the relative deviation.}
\label{div1}
\end{center}
\end{figure}

In Fig.\ref{div1}, we quantitatively compare the FDTD generated $E_y$ field with the analytical expression of Eq. (\ref{pulse}). The top row compares the absolute values of the fields at time $t=213.5~{\rm fs}$ for the different orders $m=0,1,2$. The bottom row shows the relative deviation defined as $\chi=|E_{\rm y}-E^{\rm ana}_{\rm y}|/E^{\rm ana}_{\rm y}$ in log-scale. In this expression, $E_{\rm y}$ is the FDTD generated field and $E^{\rm ana}_{\rm y}$ is from the analytical expression.  The relative deviation lies in the range of 0.01-7.0 percent over the full computation window. Interestingly, the error is reduced to less than 3 percent in the central region of the three main lobes which is obviously of higher interest. Since the deviation increases toward the antenna location, we expect that better agreement might be achieved using a larger annulus radius.

\begin{figure}[hbtp]
\begin{center}
\includegraphics[width=0.75\columnwidth]{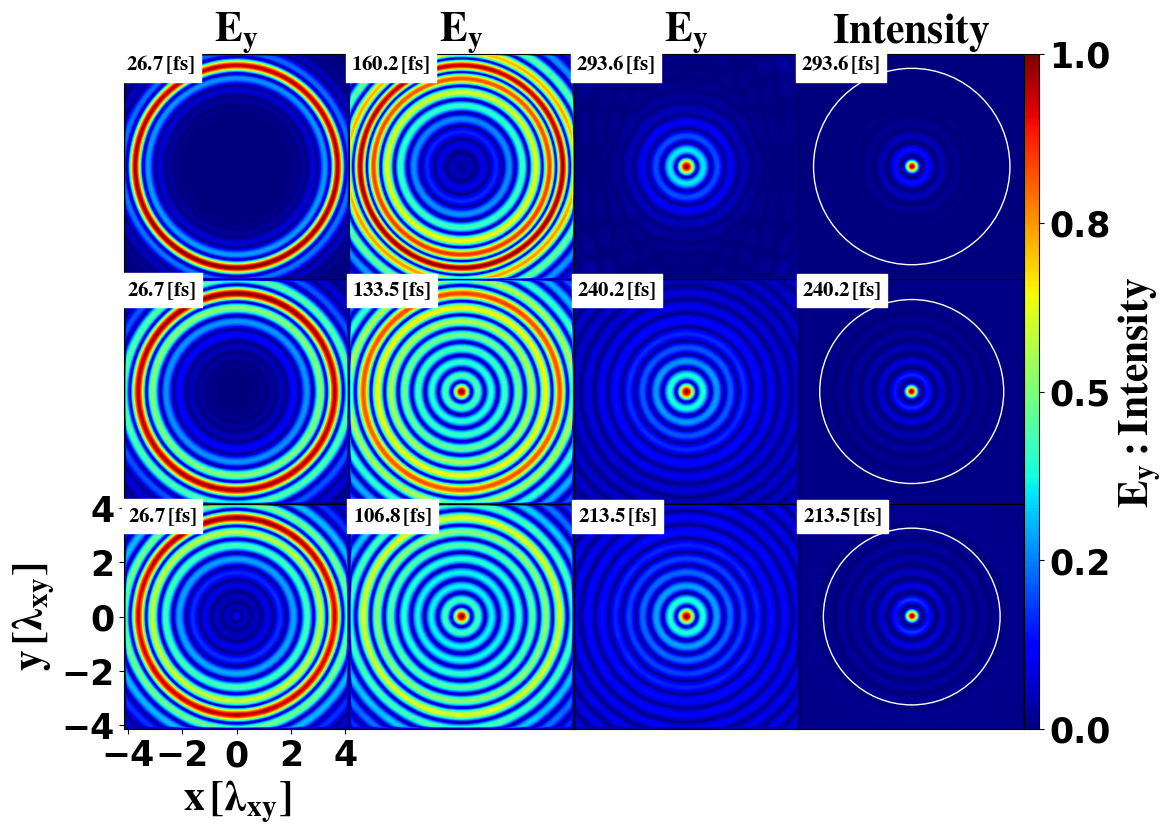}
\caption{Bessel beam with different cone angles. The amplitude of the $y$ component electric field and the intensity of generated Bessel beams are plotted on $xy$ plane at $z=0$: Run E (top row),  Run D (middle row), and Run A (bottom row). The first three columns show the time evolution of the $y$ component electric field at three different times from left to right. As the simulation box is different for each row, the corresponding time for each column is different. The rightmost column shows the intensity distribution. The white circle in the right column shows the inner radius of the  Bessel antenna.}
\label{diffang}
\end{center}
\end{figure}

\begin{figure}[hbtp]
\begin{center}
\includegraphics[width=0.75\columnwidth]{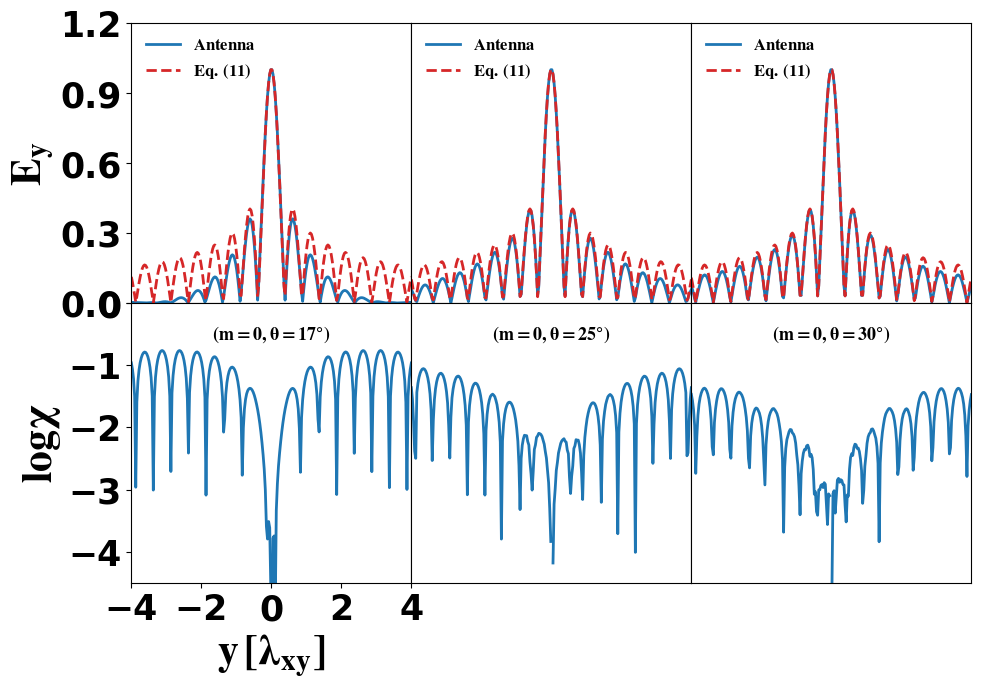}
\caption{Bessel beam with different cone angles. Comparisons between the $y$ component electric field from the Bessel antenna (blue solid line) and from the Eq. (\ref{pulse}) (red dashed line). Run E (left column),  Run D (middle column), and Run A (right column). The bottom row shows the relative deviation.}
\label{div2}
\end{center}
\end{figure}

We have also investigated the quality of the beam generation for different cone angles. The results are shown for zeroth-order Bessel beam in Fig. \ref{diffang}. The radius of the antenna was scaled according to the transverse period (see Table 1), while using an identical antenna thickness  of $1\,\lambda$. The increase in radius required longer computation time to let the pulse reach the central region. In all cases, the beams remain of high quality. Fig. \ref{div2} compares quantitatively the FDTD generated beam to the analytical one. We observe that while the error remains small - and even decreases - for lower cone angles at the center, it increases on the edges. More precisely, the number of constructed lobes decreases as the cone angle is reduced. This issue is still under investigation.

\begin{figure}[ht]
\begin{center}
\includegraphics[width=0.75\columnwidth]{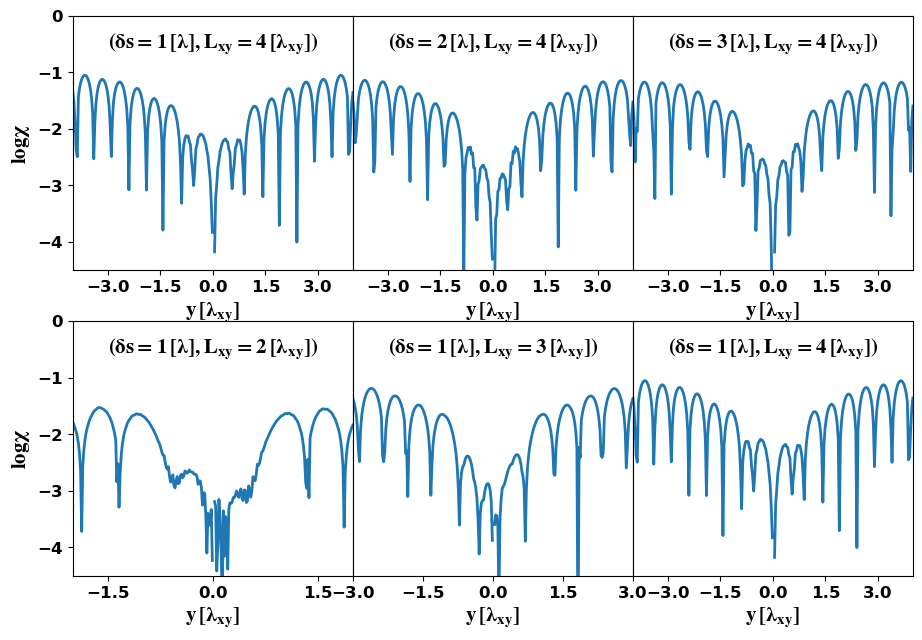}
\caption{Effects of the antenna thickness and radius on the Bessel beam. The top row shows the relative deviation for different antenna thicknesses: Run D (top left),  Run F (top middle), and Run G (top right). The bottom row shows the relative deviation for different antenna radii: Run I (bottom left),  Run H (bottom middle), and Run D (bottom right).}
\label{diffthick_diffrad}
\end{center}
\end{figure}

We have investigated the effect of antenna thickness $\delta s$ for a fixed cone angle of $\theta=25 ^{\circ}$. We show the results in Fig. \ref{diffthick_diffrad} (top row) for three different thicknesses. As one can observe on the figure, the deviation is reduced to about 1.5 percent when the antenna thickness increases. 

Finally, the radius of the Bessel antenna $R_{\rm B}$ is another important parameter since this is directly related to the computational effort. The bottom row of Fig.\ref{diffthick_diffrad} shows the deviation when  the radius of the Bessel antenna is varied from 2 to 4 transverse wavelengths $\lambda_{\rm xy}$, while maintaining the antenna thickness fixed. More lobes are obviously generated inside the antenna (with the same transverse period) for the larger box, but there is not a major difference for the three central lobes. In fact, the deviation for the three central lobes is similar in the three cases, but we note that the deviation increases for the outer lobes for the largest box. Therefore, the case of $2\lambda_{\rm xy}$ offers an excellent compromise between accuracy in the central lobes and computational effort.

We have investigated other components of the electric fields ($E_{\rm x}$, $E_{\rm y}$) as well. The relative deviation for the $E_{\rm z}$ component is similar as the deviation for $E_{\rm y}$. The $E_{\rm x}$ component which is zero in the injected solutions is four orders of magnitude smaller than the main component $E_{\rm y}$. Therefore, the electromagnetic fields generated by the Bessel antenna well satisfy Maxwell's curl equations. 
\clearpage
\section{Conclusions}\label{Conclusions}

The generation of Bessel beams in FDTD is an essential and challenging task to enable the investigation of the Bessel beam-plasma interaction with PIC codes. A common implementation is using the so-called Bessel-Gauss beam in which a Bessel beam is generated by focusing a Gaussian beam using an axicon lens, which has been demonstrated here for reference. However, due to its computational cost, we have developed an  
alternative approach that is computationally more efficient. This solution is based on using a cylindrical antenna emitting electromagnetic fields inwards and using the periodic boundary conditions along the optical axis.

We have first provided the full set of  Bessel's solutions of Maxwell's equations. The solutions have been then injected into the FDTD grid through the cylindrical antenna. We have validated our approach by comparing the field from the FDTD simulation with that from the analytical solution. We have shown that different orders of Bessel beams can be successfully generated, with small deviations with regard to the analytical solution, for different beam cone angles. Better agreement was found for the higher angles. We have investigated the effect of the different antenna parameters on the deviation and shown that thicker antenna provide a better result, while the radius of the antenna impacts negligibly on the accuracy of central lobes. The electromagnetic fields generated by the Bessel antenna well satisfy Maxwell's curl equations. In conclusion, using a computational box with a full length of two longitudinal periods and a full width of four transverse periods provides good accuracy. In comparison with the already short Bessel-Gauss beam that was generated, the computational effort has been decreased by a factor 64.
{{Despite restricted to study longitudinally periodic objects or longitudinally periodic particle distributions, we believe that our work will find applications in the field of laser-particle scattering \cite{Cui2013}, particle trapping, nonlinear plasmonics \cite{Kauranen2012} and specifically in laser-particle acceleration and laser-plasma interaction \cite{Hafizi97,Kumar2017,Faccio12,Durfee1993,Hine16}}}.

\section*{Funding}
European Research Council (ERC) (682032-PULSAR), R\'{e}gion Bourgogne Franche-Comt\'{e} and the EIPHI Graduate School (ANR-17-EURE-0002).

\section*{Acknowledgments}
We thank the EPOCH support team for help, \url {https://cfsa-pmw.warwick.ac.uk}. Numerical simulations have been performed using the M\'{e}socentre de Calcul de Franche-Comt\'{e}, PRACE Research Infrastructure resource MARCONI-KNL based at CINECA, Casalecchio di Reno, Italy, within the Project "PULSARPIC" (PRA19\_4980), and KDK computer system at Research Institute for Sustainable Humanosphere, Kyoto University.
\section*{Disclosures}
The authors declare no conflicts of interest.


\bibliography{mybib}

\end{document}